\documentclass[12pt]{article}
\usepackage{graphicx}
\usepackage{axodraw}
\usepackage{epic}
\usepackage{amssymb,amsmath}
\def\beq#1\eeq{\begin{equation}#1\end{equation}}    
\def\bea{\begin{eqnarray}}  
\def\eea{\end{eqnarray}}  
\def\bq{\begin{quote}}  
\def\eq{\end{quote}}  
\def\bi{\begin{itemize}}  
\def\ei{\end{itemize}}  
\def\be{\begin{enumerate}}  
\def\ee{\end{enumerate}}  
\def\ba{\begin{array}}  
\def\ea{\end{array}} 
\def\bi{\begin{itemize}}
\def\ei{\end{itemize}} 
\def\bc{\begin{center}}
\def\ec{\end{center}}  
\def\pa{\partial}

\def\cl{{\cal L}}

\def\rt{\sqrt{2}} 
\def\ra{\rangle}
\def\la{\langle}
\def\bi{\begin{itemize}}  
\def\ei{\end{itemize}}    
\def\dy{ \Delta y}
\def\ov{\overline}
\def\nn{\nonumber \\}

 \def\no{{\cal N}=1}

\def\k{\kappa}
\def\e{\epsilon}
\def\ov{\overline}  
\def\nn{\nonumber \\}  
\def\ch{ \left ({\Phi_1  \dots \Phi_{N-1} \over M_P^{N-1}} \right )}  
\def\ms{M_{\rm SUSY}}

\newcommand{\fref}[1]{fig. \ref{f.#1}} 
\newcommand{\eref}[1]{eq. (\ref{e.#1})} 
\newcommand{\erefn}[1]{ (\ref{e.#1})}

\newcommand{\sref}[1]{Section \ref{s.#1}}
\newcommand{\cref}[1]{Chapter \ref{c.#1}}

\newsavebox{\moose}
\sbox{\moose}{%
\begin{picture}(0,0)
  \thicklines
  \put(-60,0){\circle{35}}
  \put(60,0){\circle{35}}
  \ArrowLine(-40,0)(40,0)
\end{picture}}

\newsavebox{\site}
\sbox{\site}{%
\begin{picture}(0,0)
  \thicklines
  \put(60,0){\circle{35}}
  \ArrowLine(-40,0)(40,0)
  \ArrowLine( 80,-60)(60,-20)
  \ArrowLine( 60,-20)( 40,-60)
\end{picture}}

\begin{document}
\pagestyle{empty}
\setcounter{page}{0}
{\normalsize\sf
\rightline {hep-ph/0303155}
\rightline {IFT-03/05}
\rightline{LPT-ORSAY 03-16}
\rightline{CPHT-RR 009.0303}
\vskip 3mm
\rm\rightline{March  2003}
}

\vskip 1.0cm
\begin{center}
{\huge \bf Deconstructed $U(1)$ and \\
Supersymmetry Breaking  
         }\\
%\vskip.1cm
\vspace*{1cm}
\end{center}
 \noindent
\vskip 0.5cm
\centerline
{\sc Emilian Dudas${}^{1,2}$,  Adam  Falkowski ${}^{3}$ {\rm and} Stefan  Pokorski${}^{3}$}
\vskip 1cm
\centerline {\em  ${}^{1}$ LPT
\footnote{Unit{\'e} Mixte de Recherche du CNRS (UMR 8627).}, 
B{\^a}t. 210, Univ. Paris-Sud, F-91405 Orsay
Cedex, France}
\centerline {\em ${}^{2}$ Centre de Physique Th{\'e}orique, Ecole 
Polytechnique, F-91128, Palaiseau Cedex, France}
\centerline{\em ${}^{3}$ Institute of Theoretical Physics, Warsaw University}
\centerline{\em Ho\.za 69, 00-681 Warsaw, Poland}
\vskip.3cm
\centerline{\tt \small Emilian.Dudas@th.u-psud.fr, Adam.Falkowski@fuw.edu.pl, 
pokorski@fuw.edu.pl}
\vskip 1.5cm

\centerline{\bf Abstract}
\noindent
We discuss supersymmetry breaking induced by simultaneous presence of 
a Wilson-line type superpotential and  boundary-localized
Fayet-Iliopoulos terms in a four dimensional theory based on
deconstruction of five-dimensional abelian gauge theories on
orbifolds. Large hierarchy between the scale of supersymmetry
breaking and the fundamental scale can be generated dynamically. The
model has several potentially interesting phenomenological applications.
We also discuss the conditions that are necessary for interpreting our $U(1)^N$ model as an ultra-violet completion of some 5d theory. In particular,  the corresponding 5d theory contains Chern-Simons couplings.   

\vskip .3cm

%Last modified: 16.03.2003

%%%%%%%%%%%%%%%%%%%%%%%%%%%%%%%%%%%%%%%%%%%%%%%%%%%%%
\newpage 

\setcounter{page}{1} \pagestyle{plain}

%%%%%%%%%%%%%%%%%%%%%%%%%%%%%%%%%%%%%%%%%%%%%%%%%%%%%%%%%%%%%%%
\section{Introduction}
%%%%%%%%%%%%%%%%%%%%%%%%%%%%%%%%%%%%%%%%%%%%%%%%%%%%%%%%%%%%%%%%

Deconstructed higher dimensional gauge theories \cite{arcoge} can be
viewed as an ultraviolet (UV) completion of gauge theories \cite{faki}  in more than 4 dimensions or as a new tool for building models in four dimensions 
\cite{arcoge_twi}. 
In the latter case one does not have to insist on exact higher 
dimensional correspondence but one just explores the
possibilities offered by the basic structure of such theories
which is a product gauge symmetry containing bi-fundamental matter.
Both views on deconstructed gauge theories may provide new theoretical
insight: completing 5d theories in the UV by their deconstructed
versions may give us more rigorous calculational tools for
non-renormalizable gauge theories and may help to understand better their 
structure, whereas studying 4d gauge theories with product gauge group
in their own sake may give us the benefits usually attributed 
to extra dimensional theories but in simple 4d setting.

Deconstruction as a UV completion of 5d super-Yang-Mills theories is
interesting from the point of view of recovering ${\cal N}=2$ supersymmetry
in four dimensions \cite{csergr}. Deconstruction as a model building tool, not
necessarily with exact correspondence to a 5d theory, is interesting
as (among other reasons) it provides a mechanism for naturally 
generating hierarchical (dimensionless and dimensionful) physical 
parameters \cite{hl}. 

In this paper we consider deconstructed supersymmetric $U(1)$
gauge theories. Several aspects of such theories with unbroken
supersymmetry are discussed in ref. \cite{faniol}. Here we propose a novel
mechanism of supersymmetry breaking based on the simultaneous presence
of a Wilson-line type superpotential and boundary-localized
Fayet-Iliopoulos (FI) terms. In the present paper we explore both 
above-described aspects of deconstruction. We begin with a simple
product $U(1)$ supersymmetric model that is a self-consistent theory
in 4d. A very important role in the construction of the model is played
by the condition of mixed gauge anomaly cancellation.
The construction of the model is presented in Section 2. From the purely 
four-dimensional point of view, the model presents an interesting new
mechanism of supersymmetry breaking, where its dominant source is an
expectation value of the D-terms, similarly as in the
scenario of supersymmetry breaking with a single anomalous $U(1)$ 
\cite{fa,bidu}.

However, as we discuss in Section 3, the simple model of Section 2 has
no 5d correspondence: its continuum limit violates 5d Lorentz
invariance. It turns out that very interesting conclusions follow from 
insisting on the correct 5d correspondence, i.e. on 5d Lorentz 
invariance and ${\cal N}=2$ supersymmetry.
% This is discussed also in Section 3.
 The cancellation of mixed gauge anomalies plays again a
crucial role in that discussion. We show that the necessary extension of
the simple model of Section 2 is highly constrained. In particular,   
 the 5d continuum theory includes the Chern-Simons term.

In Section 4 we return to the simple model of Section 2 which, as we
said, is sufficient as a model of supersymmetry breaking and
calculationally simpler than the one of Section 3. Thus, in Section 4
we minimize the scalar potential and show that supersymmetry is indeed
spontaneously broken, with the scale of supersymmetry breaking
suppressed with respect to the fundamental scale by the factor $\e^N$,
where $N$ is the number of $U(1)$ gauge groups and $\e \sim 0.1$.
In the present case the scale of supersymmetry breaking is dynamically
determined by the model. Furthermore, if we embed the model into
a locally supersymmetric version, supersymmetry breaking is of a
hybrid type, with both D-term and F-term breaking but with D-term
dominating. Finally, in Section 5 we briefly discuss potential
phenomenological consequences of such a scenario for
fermion mass generation and for a solution to the supersymmetric
flavour problem.
% We also present our conclusions. 
%%%%%%%%%%%%%%%%%%%%%%%%%%%%%%%%%%%%%%%%%%%%%%%%%%%%%%%%%%%%%%%%%%%%%%%%%%%%%%%%%%%%%%%%%

\section{A simple four-dimensional model with product $U(1)$}
\label{s.m}
%%%%%%%%%%%%%%%%%%%%%%%%%%%%%%%%%%%%%%%%%%%%%%%%%%%%%%%%%

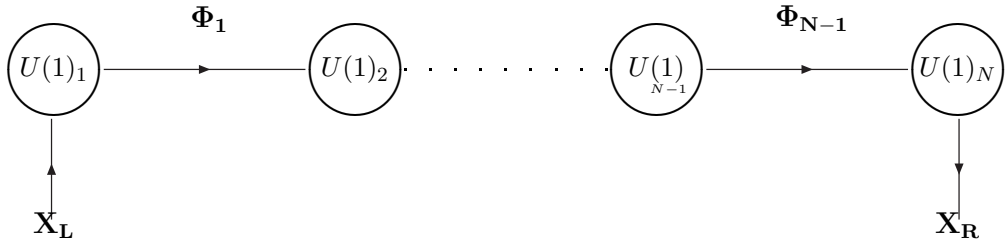
\begin{figure}[htb] 
\centering
  \scalebox{.95}{%  
\begin{picture}(-100,150)(0,0)
      \thicklines
      \put(-180,100){\usebox\moose}
      \put(60,100){{\usebox\moose}}
%\ArrowArc(135,100)(20,-115,115)
      \put(-60,100){{\makebox(0,0){\dottedline{10}(-40,0)(40,0)}}}
      \put(-240,100){\makebox(0,0){\small $U(1)_1$}}
      \put(-120,100){\makebox(0,0){\small $U(1)_2$}}
      \put(0   ,100){\makebox(0,0){\small $U(1)$}}
      \put(5   ,92){\makebox(0,0){\tiny ${}_{N-1}$}}
      \put( 120,100){\makebox(0,0){\small $U(1)_N$}}
      \ArrowLine(-240,40)(-240,80)                                                                                                                                                                        
      \ArrowLine(120,80)(120,40)
      \put(-180,120){\makebox(0,0){ $\bf \Phi_1$}}
      \put(  60,120){\makebox(0,0){ $ \bf \Phi_{N-1}$}}
 %     \put( 160,120){\makebox(0,0){\bf $\Phi_{N}$}}
       \put(-240,38){\makebox(0,0){ $\bf X_L$ }}        
  \put(120,38){\makebox(0,0){$\bf X_R$}}   
    \end{picture}
} 
\caption{The quiver diagram of the model}
\label{f.qd}
\end{figure}
 
We consider supersymmetric theories with a product $U(1)$ gauge group. 
 The setup involves the product  gauge group $U(1)_1 \times \dots U(1)_N \equiv U(1)^N$ and $N-1$ chiral multiplets (links) $\Phi_p$  charged under the neighboring groups. The quiver diagram for our model is given in  \fref{qd}. The link $\Phi_p$ has charge $Q$ under  $U(1)_p$ and $-Q$ under $U(1)_{p+1}$. The first and the $N$-th group are not linked. Furthermore, at the boundaries  of the group product space we add chiral multiplets: $X_L$ with charge $-Q$ under $U(1)_1$   and $X_R$   with charge $Q$ under $U(1)_N$. In the following we normalize $Q=1$. We shall also assume that some matter chiral multiplets (including the MSSM matter) live at the boundaries, that is they transform under $U(1)_1$ or $U(1)_N$.
In general, we consider the situation when ${\rm Tr \ Q_1} \not=0 $,  
${\rm Tr \ Q_N} \not=0 $.

Note that, for $N$ large enough, we cannot write any renormalizable potential for the links. Still, the  symmetries of the theory allow for a non-renormalizable superpotential of the Wilson-line type,  
\beq
\label{e.dfpsuperpotential}
W = M_P X_L \ch  X_R \ , 
\eeq 
where we identified the fundamental scale with the Planck scale. 
This superpotential will be the source of supersymmetry breaking.

The model as it stands is inconsistent as the $U(1)$ gauge symmetries are anomalous. There are two kinds of anomalies. The first are mixed anomalies of the neighboring groups, that are  produced by the presence of the links. The second are the anomalies of the first and the $N$-th group produced by the matter multiplets living at the boundaries. 
It is well known (see, e.g., ref. \cite{weinberg}) that the structure of anomalies depends on the  renormalization conditions imposed on the divergences of the currents. Let us define the correlator of three gauge currents
$\Gamma_{pqr}^{\mu \nu \rho} (x,y,z) = \langle 0 | T {j_p^{\mu}(x)
j_q^{\nu}(y) j_r^{\rho}(z)} |0  \rangle \ $, 
where $j_p^{\mu}$ is the chiral current coupling to the $p$-th gauge
field.  The only non-vanishing divergences relevant for the mixed anomalies 
are those involving $\Gamma_{p,p+1,p+1}^{\mu \nu \rho}$ and $\Gamma_{p+1,p,p}^{\mu \nu \rho}$.
%There are actually two consistent regularizations (see, e.g.,\cite{weinberg}).
One possible regularization is such that the anomaly is placed in the current appearing only once in the correlator.  In such regularization the gauge variation of the model  is given by:
\beq
\label{e.mixedanomaly1}
\delta \cl_{an} =  - {i \over 4 \pi^2} \sum_p \int d^2 \theta \
\Lambda_p \ \left  ( W_{p+1}^{\alpha} W_{\alpha, p+1}
- W_{p-1}^{\alpha} W_{\alpha , p-1} \right ) + {\rm h.c} \ 
%\label{a13}
\eeq
Here $\Lambda_p$ are the infinitesimal parameters of the $U(1)^N$
gauge transformations   written in the superfield formalism. 
These anomalies can be canceled by  the Green-Schwarz mechanism. To
this end the theory should contain 
superfields $M_k$ coupling to the gauge fields via the gauge kinetic functions,
%\beq
$ f_p \ = \ {1 \over g_p^2} + \sum_k s_{pk} M_k$. 
%\ . \label{a2} \eeq
Under the gauge transformations, the fields $M_k$ must transform non-linearly
%\bea&& 
$V_p \rightarrow V_p +  i(\ov \Lambda_p - {\Lambda}_p)$ ,
%\nonumber \\ && 
$M_k \rightarrow M_k +  2 i  \sum_k \epsilon_{kp} \Lambda_p \ ,$. 
%\label{a3}\eea
  To cancel  the  anomalies, the Green-Schwarz condition must be satisfied, 
\beq
C_{pq} =   \ {1 \over 2} \sum_k s_{pk} \ \epsilon_{kq} \, ,
\eeq
% \label{a4} \eeq
where the $N \times N $ anomaly matrix $C$, defined as $ C_{pq} \equiv \ {1 \over 4 \pi^2} {\rm Tr} \ (Q_ p Q_q^2)$, reads:  
\beq
\label{e.anomalymatrix}
C_{pq}  = {1 \over 4 \pi^2} \ ( \delta_{p,q-1} - \delta_{p,q+1}  
+  {\rm Tr} (Q_1)^3 \delta_{p,1}\delta_{q,1}  
+  {\rm Tr} (Q_N)^3 \delta_{p,N}\delta_{q,N})  \ . 
%\label{a1} 
\eeq

In the setup at hand, there is a natural solution to these
constraints. Note that the role of the moduli $M_k$ can be played
by the links. In the superspace 
language, the links transform as 
$\Phi_p \to e^{2 i \Lambda_p}\Phi_p e^{-2 i \Lambda_{p+1}}$.
Hence we can define an object  that transforms non-linearly, 
\beq
 \log(\Phi_{p-1} \Phi_p)\ \rightarrow \ \log(\Phi_{p-1} \Phi_p) + 2
 {i} \ (\Lambda_{p-1} -\Lambda_{p+1}) \  
\label{a05} 
\eeq
If the model were described by a periodic quiver diagram, only mixed anomalies would be present and these can be canceled by using the links only. However in the present case the group product space has boundaries where anomalies can appear. 
 Therefore  we introduce  new superfields, $S_L$ and $S_R$, transforming as
\bea
S_L \ &\rightarrow& \ S_L + {i} M_P  \ ({\rm Tr} (Q_1^3)-1)  \Lambda_{1}
\nn
 S_R \ &\rightarrow& \ S_R +  {i} M_P\ ( {\rm Tr} (Q_N^3)+1) \Lambda_{N} \  
\label{a6} 
\eea
Then anomalies \erefn{mixedanomaly1} are canceled  provided  the gauge kinetic function are chosen as:  
\bea
 f_1 &=&  {1 \over g^2}  -  {1 \over 2 \pi^2} \log \left
   ({\Phi_{1}\over v } \right )  +  
{1 \over  \pi^2 M_P}  S_L \delta_{p,1}
\nn
 f_p \ &=&  {1 \over g^2}   -  {1 \over 2 \pi^2} \log 
\left ({ \Phi_{p-1} \Phi_p \over v^2} \right ) \hspace{2cm} p=2 \dots N-1  
\nn 
f_N &=&  {1 \over g^2}  -  {1 \over 2 \pi^2} \log \left
  ({\Phi_{N-1}\over v} \right ) +   
{1 \over \pi^2 M_P} S_R \delta_{p,N} \, ,
 \ . \label{a7} 
\eea
where $v$ is an arbitrary scale which, for convenience, is chosen
equal to the links vevs. Also, we 
have set all the $U(1)$ gauge couplings to be equal for simplicity. 
An interesting  feature of this  theory is that  all the gauge couplings vanish in the unbroken  phase  $\langle \Phi_p \rangle =0$. Therefore, the model in the UV is a free theory. 
% As ${1 \over g_p^2} = \la f_p \ra$, all the gauge couplings vanish 
% in this phase. Therefore, the theory in the UV is a free theory. 

We assume the Kahler potential of the form: 
\bea&
{ \cal K} = 
\sum_{p=1}^{N-1} |\Phi_p|^2 e^{-2 V_p + 2 V_{p+1}}
 + |X_L|^2  e^{2 V_{1}} +  |X_R|^2  e^{-2 V_{N}} 
&\nn&
+   {1 \over 2} [S_L+\ov S_L  +  M_P( {\rm Tr} (Q_1)^3-1) V_{1}]^2
 +  {1 \over 2} [S_R+\ov S_R  +  M_P( {\rm Tr} (Q_N)^3+1) V_{N}]^2 \, .
\eea
 The links have a minimal  kinetic term.  
The presence of vector multiplets in the kinetic term for $S_L$ and $S_R$ makes the Kahler potential  gauge invariant. It also  generates FI terms at the boundaries of the group product space, that  are  proportional to the vacuum expectation values of $S_L$ and $S_R$, 
\bea
\xi_1 &=& M_P \ ({\rm Tr} (Q_1)^3-1) \ \la S_L+\ov S_L \ra  
\nn 
\xi_N &=& M_P  \ ({\rm Tr} (Q_N)^3+1) \ \la S_R+\ov S_R \ra   
\eea
 We will not construct an explicit superpotential that gives vevs to
 $S_L$ and $S_R$.  Instead 
we  simply assume that their vevs are such that 
$\xi_1 \sim \xi_N \sim \e M_P$, with $\e \sim 0.1$ and in the
following we will ignore their dynamics. 

In Section 4 we show that the model we have constructed provides an interesting new mechanism of supersymmetry breaking. However, first, in Section 3 we discuss the model from the point of view of the correspondence with 5d gauge theories.

%%%%%%%%%%%%%%%%%%%%%%%%%%%%%%%%%%%%%%%%%%%%%%%%%%%%%%%%%%%%%%%
\section{Anomalies and consistent five-dimensional limit}
\label{s.fdi}
%%%%%%%%%%%%%%%%%%%%%%%%%%%%%%%%%%%%%%%%%%%%%%%%%%%%%%%%%%%%%%%

An interesting point we want to discuss is the construction of a model
with a consistent 5d limit and the role of anomaly
cancellation in the 4d  deconstructed version.  In order  
to cancel the mixed anomalies we used the Green-Schwarz mechanism, and
a vital role was played by the 
links coupled to the gauge fields via the gauge kinetic
function. Recall  \cite{csergr} that the links 
contain the degrees of freedom that are translated to the fifth
component of the 5d gauge field. More 
precisely, the links can be represented as 
 $\Phi_p = {v \over \rt}e^{(\Sigma_p + i G_p)/ v}$ and $G_p$ is  what matches
 $A_5$ (while $\Sigma_p$ matches the scalar singlet  of the 5d vector multiplet).    Thus  a natural  candidate 
for a 5d match to our Green-Schwarz mechanism  is the 5d Chern-Simons term,   
\beq
\label{e.dfpcs}
\cl_{CS} = {1 \over 24 \pi^2}
\e_{\alpha\beta\gamma\delta\epsilon} \left [
 A_\alpha \pa_\beta A_\gamma \pa_\delta A_\epsilon \right] 
={1 \over 24 \pi^2} \e_{\mu\nu\rho\sigma}\left [
3 A_5  \pa_\mu A_{\nu}\pa_\rho A_{\sigma} 
 - 2   A_{\mu}\pa_\nu A_{\rho} \pa_5 A_{\sigma}
\right ] 
\eeq 
One can check that the terms we have proposed in \sref{m} do not have
a 5d invariant continuum limit and 
therefore cannot correspond to the 5d Chern-Simons term.
5d Lorentz invariance in the continuum limit must be imposed as
 an additional  constraint. 
The remainder of this section is devoted to finding a 4d action  that
can cancel the mixed anomalies and, 
at the same time,  match the Chern-Simons term in the continuum limit.

Let us consider for simplicity the simpler case with a  closed
(periodic) quiver diagram. The model has only mixed anomalies and,
using the same anomaly renormalization scheme as in Section 2,
\eref{anomalymatrix} and eq. (\ref{a7}) simplify to $C_{pq} = {1 \over
4 \pi^2}(\delta_{p,q+1} - \delta_{p,q-1})$ and  
\beq
\label{e.gaugekinetic}
f_p \ =  {1 \over g^2}  + {1 \over 2 \pi^2} \chi_p \ , 
%\label{fp}
\eeq
where $\chi_p = - \ln (\Phi_{p-1}\Phi_p/v^2)$. 
In the infrared we want to recover a 5d ${\cal N}=1$ supersymmetric theory
compactified on a circle. From a four-dimensional viewpoint, this
should be a theory with two supersymmetries ${\cal N}=2$. It is known that in 5d that the gauge couplings can be functions of
scalars $\Sigma_p$ in the vector multiplets , which is indeed consistent with our identification of the gauge kinetic function. 
%(\ref{a04}), (note that $\Sigma^p = \sqrt{2} \ {\rmRe \chi}^p$).
 Moreover, the couplings of the
vector multiplet are completely specified by a real function, the prepotential $
{\cal F}(\Sigma_p)$, which is a polynomial function at most trilinear in the
scalar fields $\Sigma_p$ (for a review of 5d abelian supersymmetric theories, see e.g. \cite{ims}). For example, the gauge couplings $\tau_{pq} (X)
F_{MN}^p F^{MN,q}$ are provided by
$\tau_{pq} = \partial_p \partial_q {\cal F} (\Sigma) \ \equiv  {\cal F}_{pq}$. 
In four dimensions (for a review, see e.g. \cite{vp}),  the gauge kinetic function  becomes a holomorphic function of the superfields $\chi^p$ and we need to substitute $\Sigma_p \to {v \over 4}(\chi_p + \bar \chi_p)$.  The effective lagrangian of vector multiplets is, by using ${\cal N}=1$ language, given as: 
\beq
{\cal L} \ = \ {1 \over 2} \int d^2 \theta \ \sum_{pq} {\cal F}_{pq}
\ W^{\alpha ,p} W_{\alpha}^q + {\rm h.c.} + 
\int d^4 \theta \ {\cal K} (\chi^p, {\bar \chi_q})  
\ , \label{a08}
\eeq
where the Kahler potential is given in terms of the prepotential by 
\beq
 {\cal K} (\chi^p, {\bar \chi}_p) \ = \sum_p \ ( {\cal {\bar F}}_p \chi^p
 + {\cal F}^p {\bar \chi}_p )  \ . \label{a8}
\eeq
By matching with \eref{gaugekinetic} we find that the 5d  prepotential and the derived 4d Kahler potential of this theory are
\bea
{\cal F}(\Sigma_p) &=& {1 \over 2 g^2} \sum_p \Sigma_p^2 - {1 \over 6 \pi^2} \sum_p \Sigma_p^3 \ , \nonumber \\
{\cal K}(\chi^p, {\bar \chi}_p) &=& {v^2 \over 16 g^2} \sum_p (\chi_p + {\bar \chi}_p + 2 V_{p-1}- 2 V_{p+1})^2 + {v^2 \over 384 \pi^2 } \sum_p
(\chi_p + {\bar \chi}_p + 2 V_{p-1} - 2 V_{p+1})^3 \ .  
 \label{a10} 
\nn
\eea
 However the action is not 5d invariant in the continuum limit, as eq. (\ref{a10}) clearly does not yield the last term of the 5d Chern-Simons
couplings (\ref{e.dfpcs}), containing  $\pa_5 A_\mu$. 

Interestingly enough, there is another consistent regularization of
gauge anomalies, which is compatible with 5d Lorentz invariance. 
In this regularization, the anomalous divergences are placed symmetrically in each current in  $\Gamma^{\mu\nu\rho}_{p,p,p+1}$ and $\Gamma^{\mu\nu\rho}_{p+1, p,p}$. 
The anomalous variation of the action is then  equal to
\beq
\label{e.symmetricanomaly}
\delta \cl_{an} =  - {i \over 12 \pi^2} \sum_p \int d^2 \theta \
\Lambda_p \
\left  ( W_{p+1}^{\alpha} W_{\alpha, p+1}
- W_{p-1}^{\alpha} W_{\alpha , p-1}
 - 2 W_{p}^{\alpha} W_{\alpha, p+1} 
 + 2 W_{p}^{\alpha} W_{\alpha,p-1} \right ) + {\rm h.c} \ , 
\eeq
whereas the anomalous couplings
%(\ref{a5}) 
present in the lagrangian (\ref{a10}) can account only for the
first two terms in \eref{symmetricanomaly}. 

In this symmetric regularization, a  Wess-Zumino term is needed and it is 
naturally selected to be
\beq
\cl_{WZ} = - {1 \over 12 \pi^2} \int d^4 \theta \ \left [ 
(V_{p+1}-V_{p-1}) D_\alpha V_{p} - V_{p} (D^{\alpha} V_{p+1} -D^{\alpha}
V_{p-1}) \right] W_{\alpha ,p} \ + \ {\rm h.c.} \ , \label{a15} 
\eeq 
whose variation under gauge transformations is
\beq
\delta \cl_{WZ} =  - {i \over 6 \pi^2} \sum_p \int d^2 \theta \
\Lambda_p \
\left  ( W_{p-1}^{\alpha} W_{\alpha, p-1}
- W_{p+1}^{\alpha} W_{\alpha , p+1}
-  W_{p}^{\alpha} W_{\alpha, p+1} 
+ W_{p-1}^{\alpha} W_{\alpha,p} \right ) + {\rm h.c} \  \label{a16}
\eeq
and which, combined with the gauge variation coming from \eref{gaugekinetic}, exactly
cancel the anomalous one-loop variation \eref{symmetricanomaly}.  

Using the dictionary $A_{\mu,p} \to A_\mu(y_p)$, $(A_{\mu,p+1} -
A_{\mu,p})/\dy \to \pa_5 A_\mu(y_p)$,  
$G_p \to A_5(y_p)$ with the lattice spacing $\dy = (v)^{-1}$, it is straightforward
to check that the full Kahler potential (\ref{a10}) supplemented by 
(\ref{a15}) is actually the deconstructed
version of the Chern-Simons one discussed in \cite{agw} and therefore
in the continuum limit, we indeed recover in the IR a 5d
supersymmetric theory. The manifestly supersymmetric massless and
massive vector multiplets are in  ${\cal N}=1$ language $(V_p,
\chi_p)$.   
On the circle,  the Chern-Simons term does not play any role in anomaly 
cancellation as its variation is a total derivative. However,
interestingly enough, in our deconstruction model this term (in its
supersymmetric from) is  present in
order to cancel mixed gauged anomalies.

%%%%%%%%%%%%%%%%%%%%%%%%%%%%%%%%%%%%%%%%%%%%%%%%%%%%%%%%%%%%%%%%%%%%%%%%%%
\section{Supersymmetry breaking}
\label{s.sb}
%%%%%%%%%%%%%%%%%%%%%%%%%%%%%%%%%%%%%%%%%%%%%%%%%%%%%%%%%%%%%%%%%%%%%%%%%%

In this section we discuss  supersymmetry breaking triggered
by the Wilson-type superpotential (\ref{e.dfpsuperpotential}). 
The model with a consistent 5d limit considered in Section 3,  on the
orbifold, is hard to analyse. We return therefore to
the simple model of Section 2 and study supersymmetry breaking.
 
The D-term potential in the model of Section 2 takes the form:
\bea
&V_D = {1 \over 2} 
\left[  ({\rm Re \ f_1})^{-1} (|\Phi_1|^2 - |X_L|^2 +\xi_1)^2 
+ ({\rm Re \ f_2})^{-1} (|\Phi_2|^2 - |\Phi_1|^2)^2 
  + \dots  
\right . &\nn& \left .  
 + ({\rm Re \ f_{N-1}})^{-1} (|\Phi_{N-1}|^2 - |\Phi_{N-2}|^2)^2  
+ ({\rm Re \ f_N})^{-1} (-|\Phi_{N-1}|^2 + |X_R|^2 +\xi_N)^2  \right ] \, .
&\eea      

Since the F-term potential is suppressed by powers of $M_P$, the
D-term dominates the scalar 
potential and, in the zeroth order approximation, the vacuum adjusts itself
to minimize it. Depending on 
values and signs of the FI terms various patterns of gauge symmetry  and supersymmetry breaking may occur. 
 %One possibility is the so called deconstruction phase when all link-Higgs fields acquire vevs while $X_L$ and $X_R$ do not, which 
%happens for the FI terms satisfying $\xi_N >0$ and $\xi_1 =
%-\xi_N$. In such case the product  group is broken down do the diagonal subgroup and, below the  scale $v$, the spectrum and  interactions of such model are similar to those of the 5d supersymmetric $U(1)$ gauge theory 
%compactified on $S_1/Z_2$ \cite{faniol}.
 Here we are  interested in
the situation when the product 
group is entirely broken, which happens for  
\beq \xi_N > 0 \hspace{2cm} \xi_1 > -\xi_N \, , \eeq 
 which we assume from now on.

In the zeroth order approximation,  ignoring the contributions from the F-term  potential and from the non-trivial gauge kinetic term,  the D-term potential possesses a vacuum solution with  a flat direction  parametrized by the  vev of $X_R$, 
% $ t = a \langle X_R^2 \rangle$
\bea
\label{e.fd}
\langle |\Phi_p|^2\rangle  &=&  \xi_N + \langle |X_R|^2 \rangle \nn
 \langle |X_L|^2 \rangle &=&  \xi_N + \xi_1 + \langle |X_R|^2 \rangle 
\eea  
In this background the product gauge symmetry is entirely
 broken. There is one massless chiral 
multiplet (for $\langle |X_R|^2 \rangle =0$ it is just $X_R$). The
 remaining degrees of freedom form a tower of 
gauge multiplets with masses starting at  $m^2 \sim \xi/N$. 
Supersymmetry is unbroken at this order.  

Now we include  the effects of the F-term potential. 
One can easily see that its addition  lifts  the flat direction and
(for a globally supersymmetric 
scalar potential) sets the minimum at  $\la  X_R^2 \ra =0$ (up to corrections suppressed by $(1/M_P)^N$). In such case, effectively, the scalar 
potential is augmented  only by  $|{\pa W \over \pa X_R}|^2$. But
since this operator originates 
from a non-renormalizable superpotential it is only a small
perturbation to the zeroth order 
supersymmetric solution. The vacuum shift is  suppressed by  the small parameter $\k$ defined as 
\beq
\k^2 = {1 \over g^2}({\xi_N \over M_P^2})^{N-2} \approx \e^{2N-4} \,.
\eeq    
We shall solve the equations of motion  to the lowest non-trivial order in $\k^2$.          
We expand the links around the zeroth-order  vacuum solution, 
\beq
|\la \Phi_p \ra|^2 =  \xi_N  + a_p \ , \
|\la X_L \ra|^2 = \xi_1 +\xi_N   + a_0 \ , \ 
|\la X_R \ra|^2 =  a_N \ , 
\eeq 
 where $a_p$ are of  order $\k^2$. One can check that effects of the non-trivial gauge kinetic function appear only at order $\k^4$. To order $\k^2$ the equations of motion read:   
\bea
\label{e.dfpequationslinearized}
  a_0 - a_1 +  \k^2 \xi_N  & = 0 \, ,
\nn 
 -a_{p-1}+ 2 a_p - a_{p+1} +  \k^2 (\xi_N + \xi_1)  
&= 0 & \;\;\; \;\; p = 1 \dots N-1 \ . 
\eea
 We encounter a difference equation of the form: $-a_{p-1}+ 2 a_p -
 a_{p+1} + X = 0$,  with 
$X = \k^2  (\xi_1 + \xi_N)$. The general solution is given by $a_p = A
 + B p + {1 \over 2} X p^2$, 
where $A$ and $B$ are arbitrary constants.  The first of
 \eref{dfpequationslinearized} acts as a 
`boundary condition' for the difference equation, $a_{-1} = a_0 +
 \kappa^2 \xi_1$. This allows to 
determine the constant $B=X/2 - \kappa^2 \xi_1$.  The constant $A$ is not determined and so  the flat direction persists at the order $\k^2$. 
However, the value of $A$  is not important in what follows, in particular, the supersymmetry breaking parameters do not depend on $A$.  
Hence  we find  that the vacuum solution to first order in $\k^2$ is given by 
\bea
\la |\Phi_p|^2 \ra &=&  \xi_N +   
 A + {1 \over 2} \k^2 \left ( p (\xi_N - \xi_1)+  p^2 (\xi_N + \xi_1) \right ) \nn
\la   |X_L|^2\ra &=& \xi_1+ \xi_N + A 
\nn 
 \la  |X_R|^2 \ra &=& A + {1 \over 2} \k^2 \left ( N (\xi_N - \xi_1)+  N^2 (\xi_N + \xi_1) \right ) 
\eea    
In this shifted vacuum supersymmetry is broken and  the expectation values of the D-terms  are: 
\bea
D_p &=& \k^2 [  p(\xi_1 + \xi_N) - \xi_1 ] \,. 
\eea 
There is also an F-term acquiring vev:
\beq
F_{X_R} = {\pa W \over \pa X_R} =  \k \sqrt{\xi_N (\xi_1 + \xi_N) \over a}
\, .
\eeq 
Note that all the D-terms are positive. The consequence of this fact
 is that  the matter  we assumed 
to be present at the boundaries  has to have non-negative  $U(1)_1$ or
 $U(1)_N$ charges. Otherwise, 
the scalars of a negatively charged multiplet would get a tachyonic
mass and render the model unstable.

The most interesting point in this construction is that, in  a natural
way,  the supersymmetry 
breaking scale is suppressed with respect to the fundamental scale $M_P$. 
Recall that $\k = \e^{N-2}$, $\xi = \e^2 M_P^2$. Defining the
supersymmetry breaking scale as the 
scale of the D-term of the first group,  $\ms^2 = D_1$,  we get:
\beq
\ms = \e^{N-1} M_P \, .
\eeq
For $\e \sim 0.1$, even for a moderate number of replications, say $N
\sim 10$,  it is easy to 
generate a huge hierarchy between the fundamental and the supersymmetry breaking scale.
Hence the `desert' above the TeV scale can simply be a consequence of
the existence of a product $U(1)$ group at some high energy scale.   
The origin  of the hierarchy is  the fact that supersymmetry breaking
is triggered by a non-local, Wilson type  object - the superpotential
of \eref{dfpsuperpotential}. 
Thus we expect the hierarchy is not particularly sensitive to the technical assumptions we have made. 

This picture is  slightly modified  when the model is embedded in
supergravity. 
The superpotential we assume here has the form
$ W_{\tt SUGRA} = \hat W(S_L, S_R) + W(\Phi_p)$, 
where $W(\Phi_p)$ is the same as in the globally supersymmetric case, see \eref{dfpsuperpotential}. 
We do not specify the  precise form of $\hat W$ but simply assume that
it  stabilizes  
$S_L, S_R$ (in the following denoted collectively as $S$) at the value close to the fundamental scale, 
$\la S + S^\dagger \ra \approx M_P$, and that 
$\la M_P {\pa \hat W \over \pa S} \ra  \approx \la \hat W \ra$. Then,
to the leading order in 
the $|\xi_N/M_P^2|$ expansion  the scalar potential takes the form         
\beq
\label{e.asp}
 V  \approx  V_{\tt GLOBAL}  + (N-2) {\la \hat W \ra \over M_P^2} (W +
 W^\dagger) - 2 { \la \hat W \ra^2 \over M_P^2} 
\eeq
To avoid a large cosmological constant,  $\la \hat W \ra$ should
 cancel the positive vacuum energy generated by the globally
 supersymmetric part of the potential $V_{\tt GLOBAL}$. The latter is
 dominated by  the F-term $F_{X_R}$,  thus we need $F_S \approx
 F_{X_R}$. The gravitino mass can be estimated to be  
$m_{3/2}  = { \la \hat W \ra / M_P^2} \approx  {F_{X_R} / M_P}$.

It is interesting to investigate what is the 
higher-dimensional theory interpretation of the  model of supersymmetry 
breaking  we presented.
 Let us consider  a  5d supersymmetric $U(1)$ gauge theory
compactified on the orbifold 
$S_1/Z_2$ \cite{csergr,faniol}. The chiral multiplets $X_L$ and $X_R$
reside at the two different fixed 
points - respectively at $x_5 =0 $ and $x_5 = \pi R$. Moreover, there
are brane localized  FI terms, 
$S_5 \sim \xi_1 \delta (x_5) +\xi_N \delta (x_5 - \pi R)$. 
When the fifth dimension is integrated out one finds a broken 4d $\no$
$U(1)$ theory with two chiral 
multiplets $X_L$ and $X_R$. At the tree-level $X_L$ and $X_R$ do not couple to each
other as they originate from  two 
sequestered branes. However, integrating out heavy gauge bosons with
masses of the order of the cut-off 
$\Lambda$  will induce a tiny coupling and one expects $W \sim \Lambda
\ e^{-\pi R \Lambda} X_L X_R$. We could view this
extra-dimensional model as a construction 
justifying the smallness of the holomorphic mass term  for the
$X_{L,R}$ fields. Rewriting the cut-off as 
$\Lambda =N /R$, where $N$ is the number of the KK modes below the cut-off we get
$ \ms \sim  \Lambda \ e^{-\pi N}$.   
This should be confronted with  $\ms \sim  M_P \kappa$ where 
$\k \sim \e^N$. In both cases the supersymmetry breaking is
controlled by a moderately small parameter 
raised to the power $N$ - number of heavy modes in the theory.

%%%%%%%%%%%%%%%%%%%%%%%%%%%%%%%%%%%%%%%%%%%%%%%%%%%%%%%%%%%%5
\section{Phenomenological consequences}
\label{s.pa}
%%%%%%%%%%%%%%%%%%%%%%%%%%%%%%%%%%%%%%%%%%%%%%%%%%%%%%%%%%%%%

In the Section 4 we showed that the scenario with replicated
$U(1)$s and a Wilson-type 
superpotential leads to supersymmetry breaking, whose scale is
naturally much lower than the fundamental 
scale. Both D-terms and F-terms expectation values are
non-vanishing. In this section we discuss 
the consequences of such hybrid scenario for low energy phenomenology.   
   
Assuming that both FI terms are of the same order $\xi$  all the
scales of the low energy 
lagrangian are determined in terms of $M_P$, $\k$ and  $\e = \sqrt
{|\xi| \over M_P^2}$. We define 
the supersymmetry breaking scale  $\ms^2$ as the magnitude of the
D-term of the first group. Orders of 
magnitude of parameters relevant for low energy phenomenology are 
\bea
M_{{\rm SUSY}}& \approx &  M_P \k \e \ , \ 
\nn
D_p &\approx& p \ms^2 \ , \ \nn
F_{X_R},\; F_S  &\approx& M_{{\rm SUSY}} M_P \e \ , \ 
%F_{X_L} \approx  M_{{\rm SUSY}} M_P \e^2 
\nn \la X_L\ra,\; \la \Phi_p \ra   &\approx& M_P \e \ . \
%\la X_R\ra  \approx M_P \e^2 \nn
\eea
 
The pattern of soft masses depends on how MSSM fields are embedded in the model.  A matter multiplet $P$ charged under $U_p(1)$ receives soft scalar mass terms from the appropriate $D$ term,
\beq
m_Q^2 = q_p \ D_p \ \approx \ q_p \ p \ \ms^2 \ ,
\eeq
where $q_p$ is the charge of $Q$ under $U(1)_p$. 
For multiplets neutral under $U(1)$ the supersymmetry breaking is transmitted via Kahler potential, e.g.  
\begin{equation}
\int d^4 \theta \ { X_R^\dagger X_R  \over M_P^2} \ Q^\dagger Q \ \to
\ m_Q^2 \ \sim  \ {F_{X_R}^2 \over M_P^2} \ Q^\dagger Q \ \approx \
\ms^2 \ \e^2 
\end{equation}
and  these soft masses pick up an additional factor $\e$.

 Soft Majorana gaugino masses are mediated by the F-term of $S$  and are  of the same order of magnitude as neutral scalars, 
\beq
\int d^2 \theta \ {S \over M_P} W_\alpha W_\alpha \ \to  \ m_\lambda
\ \approx \ {F_S \over M_P} \ \approx \ \ms \e \ .     
\eeq
This is also the order of magnitude of the gravitino mass. 

Thus we see that the MSSM spectroscopy could exhibit two different
scale, $\ms$ and $\e \ms $, that 
differ by an order of magnitude. The possibility of such a  splitting
among superpartner masses is 
very advantageous from the phenomenological point of view. If the
first two generation squark and 
leptons are much heavier than those of the third generation, then one
can reconcile the naturalness 
bounds with the constraints arising from flavour changing neutral current processes.     

 An interesting thing about supersymmetry breaking by anomalous $U(1)$
 is that all the necessary 
ingredients to implement the Froggatt-Nielsen mechanism \cite{frni} are already at
 hand. The MSSM matter fields 
need to have positive $U(1)$ charges (otherwise they would acquire
 negative mass squares). In  Yukawa 
interactions  the excess charge has to be compensated by coupling to
 the appropriate power of  the 
negatively charged field $X_L$. If $U(1)$ acts differently on the
 three generations, various Yukawa 
interactions are suppressed by powers of the  parameter $\e =
 \sqrt{\xi/M_P^2}$. Note that $\e$ has 
generically the order of magnitude of the Cabbibo angle. In the
 following we shall assume $\e \approx 0.2$

Froggatt-Nielsen mechanism works also in the product $U(1)$
case. Proceeding along the lines of 
\cite{bira,duposa} one assumes that all quarks are charged under  the
first group,  $U(1)_1$. All quark 
masses come from supersymmetric interactions (in general non-renormalizable ones) with $X_L$: 
\beq
W = \lambda^U_{ij} H_u Q_i U_j ({X_L \over M_P})^{h_u + q_i + u_j} + 
  \lambda^D_{ij} H_d Q_i D_j ({X_L \over M_P})^{h_d + q_i + d_j} \ , 
\eeq
where we denote the $U(1)_1$ charges of the higgses, left-handed
quarks and right-handed quarks by 
$h_u,h_d,q_i, u_i, d_i$, respectively.  
Various charge assignments leading to acceptable mass and mixing
patterns are  summarized in 
ref. \cite{duposa}. 

The mechanism of fermion mass generation can control also the squark mass
pattern. In the flavour basis 
 the off-diagonal entries 
in the squark mass matrix originate from the Kahler potential and are
expected to be of order $(m_F^2)_{ij} \sim  \ms^2 \e^{2 + |f_i - f_j|}$. 
However, in the fermion mass eigenstate basis the non-diagonal contributions in  the squark mass matrix can be generated also from a splitting of diagonal entries in the flavour basis. Thus, to solve the supersymmetric flavour problem one has to control also the diagonal entries \cite{duposa2}. The dominance of the D-term breaking considered in ref. \cite{bidu} and in the present paper offers such a mechanism.    

%%%%%%%%%%%%%%%%%%%%%%%%%%%%%%%%%%%%%%%%%%%%%%%%%%%%%%%%%%%%%%%
\section{Conclusions} 

In this paper we have studied supersymmetry breaking in theories with a product $U(1)$ group and bi-fundamental matter. Fayet-Iliopoulos terms at the boundary of the group product space together with the Wilson-line type superpotential trigger supersymmetry breaking. 
A very interesting thing about this  setup is that the scale of supersymmetry breaking hierarchically lower than the fundamental scale can be generated dynamically.   

Furthermore, we have shown that interesting conclusions follow from insisting on a 5d Lorentz invariant continuum limit of such theories. Cancellation of mixed anomalies requires introducing Wess-Zumino terms in the 4d theory which, in the continuum limit, match the 5d Chern-Simons terms. Hence a consistent UV completion of our model is a  5d supersymmetric U(1) theory that contains the Chern-Simons couplings. Similar conclusions are expected to hold for U(n).       

%%%%%%%%%%%%%%%%%%%%%%%%%%%%%%%%%%%%%%%%%%%%%%%%%%%%%%%%%%%

\section*{Acknowledgments}

We wish to thank Z.~Lalak, J.~Mourad and S.~Theisen for useful discussions. 
ED, AF and SP   were supported in part by the RTN European Program
HPRN-CT-2000-00148. 
SP was partially supported Polish KBN grant 2 P03B 129 24   for years 2003--2004.  AF was partially supported Polish KBN grant 2 P03B 070 23 for years 2002--2004.

%%%%%%%%%%%%%%%%%%%%%%%%%%%%%%%%%%%%%%%%%%%%%%%%%%%%%%%%%%%%%%%%%%%%%%%%%%%%%%%%%%%%%%%%%%%%%%%

\end{document}